\documentclass[aps,prb,twocolumn,groupedaddress]{revtex4-1}

\usepackage{hyperref}
\hypersetup{
    bookmarks=true,         
    unicode=true,          
    pdftoolbar=false,        
    pdfmenubar=true,        
    pdffitwindow=true,      
    pdftitle={My title},    
    pdfauthor={Author},     
    pdfsubject={Subject},   
    pdfnewwindow=true,      
    pdfkeywords={keywords}, 
    colorlinks=true,       
    linkcolor=red,          
    citecolor=green,        
    filecolor=magenta,      
    urlcolor=cyan           
}

\usepackage[all]{hypcap}
\usepackage{graphicx}
\usepackage{amsmath}

\usepackage{layouts}

\begin{document}
\title{Equation of state of paramagnetic CrN from ab initio molecular dynamics}

\author{Peter Steneteg}
\email[]{peter.steneteg@liu.se}
\author{Bj\"orn Alling}
\author{Igor A. Abrikosov}
\affiliation{Department of Physics, Chemistry and Biology (IFM), 
Link\"oping University, SE-581 83 Link\"oping, Sweden}

\date{\today}

\begin{abstract}
Equation of state for chromium nitride has been debated in the literature in connection with a proposed collapse of its bulk modulus following the pressure induced transition from the paramagnetic cubic phase to the antiferromagnetic orthorhombic phase [F. Rivadulla \emph{et al.}, Nat Mater 8, 974 (2009); B. Alling \emph{et al.}, Nat Mater 9, 283 (2010)]. Experimentally the measurements are complicated due to the low transition pressure, while theoretically the simulation of magnetic disorder represent a major challenge.
Here a \emph{first-principles} method is suggested for the calculation of thermodynamic properties of magnetic materials in their high temperature paramagnetic phase. It is based on \emph{ab-initio} molecular dynamics and simultaneous redistributions of the disordered but finite local magnetic moments. We apply this disordered local moments molecular dynamics method to the case of CrN and simulate its equation of state. In particular the debated bulk modulus is calculated in the paramagnetic cubic phase and is shown to be very similar to that of the antiferromagnetic orthorhombic CrN phase for all considered temperatures.  
\end{abstract}

\maketitle

\section{Introduction}
Chromium nitride is a material which combines practical and industrial relevance as a component in protective coatings \cite{Vetter1995, Reiter2005} with fascinating fundamental physical phenomena. The latter include a phase transition with a magnetically driven lattice distortion\cite{Fillipetti-PRL-85-5166} between an antiferromagnetic orthorhombic low temperature phase and a paramagnetic cubic high temperature phase\cite{Corliss1960-PhysRev-117-929}. The importance of strong electron correlations as well as the necessity to model the paramagnetic state using finite disordered local moments have been recently shown\cite{Herwadkar2009-PRB-79-035125,Alling2010-PRB-82-184430}. Important issues, such as the impact of the phase transition on the compressibility of the material\cite{Rivadulla2009-Nat.Mater-8-947, Alling2010-NatMater-9-283} as well as on the electrical conductivity\cite{ Bhobe2010-PRL-104-236404, Zhang2011_hopping} are still subjects of an intense discussion.

The core problem of obtaining a complete understanding of these phenomena and properties on the most fundamental level of physics arises from the difficulty of simulating the paramagnetic high-temperature phase from first principles. In this work we first discuss the methodologies that have been used in theoretical treatments of paramagnetism. Then we present a practical scheme for calculating thermodynamic properties, in particular the equation of state, of a paramagnetic material at elevated temperature merging \emph{ab initio} molecular dynamics (MD) and the disordered local moments model (DLM). This DLM-MD technique is then applied to investigate the influence of temperature and pressure on the compressibility of CrN. We show that the change of the bulk modulus of CrN upon the pressure induced phase transition is minimal, strengthening conclusions from earlier static calculations~\cite{Alling2010-NatMater-9-283, Alling2010-PRB-82-184430} which questioned its reported collapse~\cite{Rivadulla2009-Nat.Mater-8-947}. 

\section{Modeling the paramagnetic state}
\subsection{Background}
A theory that describes the finite-temperature aspects of itinerant electron magnets has to take into account the existence of local magnetic moments present above the magnetic transition temperature, the Curie temperature $T_C$ or the Ne\'el temperature $T_N$ for a ferromagnetic or an antiferromagnetic material respectively~\cite{Moriya1985}. 
At the same time, the majority of methods used for \emph{ab initio} electronic structure calculations nowadays are based on the density functional theory (DFT) in the local (local spin density, LSDA) or semi-local (generalized gradient, GGA) approximations. While they are known to give an accurate description of the ground state properties of magnetic systems~\cite{James1999},
its straightforward generalization to finite temperatures leads to quantitative, as well as qualitative errors~\cite{Gyorffy1985-JPhysF-15-1337}. Indeed, $T_C$ of transition metals are overestimated by a factor of five and there are no moments and no Curie-Weiss law above $T_C$.
A solution to this problem should in principle be sought in the physics of strongly correlated electron systems. In particular, the dynamical mean-field theory (DMFT)~\cite{Georges1996}, combined with LDA band structure calculations has been used with success for simulations of finite-temperature magnetism in Fe and Ni~\cite{Lichtenstein2001}. However, its application to the study of structural phase transition in Fe~\cite{Leonov2011} had to neglect a contribution from the lattice dynamics, because of prohibitively high computational cost and difficulties in calculating forces between atoms~\cite{Kotliar2006}.

At the same time, it is realized that LSDA calculations at zero temperature can provide valuable information for the description of the finite temperature magnetism. One way of doing this is to extract magnetic interactions in the form of exchange constants for a classical Heisenberg model~\cite{Liechtenstein1987} or magnetic ``forces'' (the first variation of the total energy for a differential rotation of a local moment)~\cite{Antropov1995, Antropov1996} from DFT calculations and to use them in statistical mechanics~\cite{Rosengaard1997, Ruban2004, Ruban2007, Alling2009, Alling2010TiCrN} or in spin-dynamics~\cite{Skubic2008, Hellsvik2008} simulations of magnetic properties at elevated temperatures. Another useful approach is given by the so-called Disordered Local Moment model, introduced by Hubbard~\cite{Hubbard1979, Hubbard1979B, Hubbard1981} and Hasegawa~\cite{Hasegawa1979, Hasegawa1980}, and combined with the LSDA-DFT by Gyorffy \emph{et al.}~\cite{Gyorffy1985-JPhysF-15-1337}. Within the DLM picture, the local magnetic moments exist in the paramagnetic state above the magnetic transition temperature, but are fully disordered. The magnetically disordered state can be described as a pseudo-alloy of equal amounts of atoms with spin up and spin down orientations of their magnetic moments, and its electronic structure and the total energy can be calculated within the conventional alloy theory using the coherent potential approximation (CPA)~\cite{Gyorffy1985-JPhysF-15-1337} or the supercell technique~\cite{Alling2010-PRB-82-184430}. The methodology is highly successful in applications to many materials systems, ranging from steels~\cite{Olsson2003, Olsson2006} to actinides~\cite{Niklasson2003}, and its generalization for the case of partial magnetic disorder can be used in simulations of structural phase transitions in a vicinity of magnetic $T_C$~\cite{Ruban:2008p094436, Ekholm2010}. Still, to the best of our knowledge all the applications of the DLM model so far neglected the effect of lattice vibrations.

On the other hand, the importance of lattice dynamics for an accurate description of thermodynamic properties of materials is well recognized by now~\cite{Walle2002, Gillan2006}. In particular, its consideration can be essential for a treatment of lattice stabilities~\cite{Mikhailushkin2007, Asker2008, Ozolins2009, Lavrentiev2010}, heat capacities~\cite{Kormann2008, Grabowski2009}, and equations of state~\cite{Isaev2011} of solids. The importance of lattice vibrations should be fully recognized in the paramagnetic state of magnetic materials, as it occurs only at elevated temperatures. However, simultaneous treatment of the magnetic disorder, inherent to the paramagnetic state, and lattice vibrations represent a truly challenging task. 

State-of-the art treatments of lattice vibrations are based either on (quasi-) harmonic calculations of the phonon dispersion relations or on molecular dynamics simulations~\cite{Martin2004}. For magnetically ordered materials these techniques can be applied straightforwardly. However, in the presence of magnetic excitations this will not work. In particular, in the paramagnetic state at high temperatures the relevant magnetic excitations are associated with spin-flips. Their characteristic time scale can be estimated by the spin decoherence time $t_{dc}$. Spin-dynamics simulations of the spin autocorrelation function in bcc Fe above $T_C$~\cite{Hellsvik2008} show that $t_{dc}$ is of the order of 20-50 fs. In materials with lower $T_C$ it should be larger by approximately a corresponding factor, because both $T_C$ and the velocity of the propagation of the local moments are related to the strength of the exchange interactions.  At the same time a typical MD run should be carried out for at least 3-5 ps, as dictated by the inverse of the Debye frequency ($\sim10^{-12}$ s). This means that magnetic configurations should change often during the MD run. Simultaneously, a typical MD time step is of order of 1 fs, which is still much smaller than $t_{dc}$. Thus, the magnetic degree of freedom is slow on the time scale relevant for the determination of temporal evolution of a particular atomic configuration, but fast on the time scale relevant for a proper exploration of the phase space of atomic configurations. Therefore, the adiabatic decoupling between magnetic and vibrational degrees of freedom cannot be applied, and they should be treated within one single framework. Similar arguments can be used to question a validity of lattice dynamics studies for paramagnetic materials based on quasi-harmonic approximation. Perhaps the most consistent approach to the analysis of spin-lattice interactions at finite temperature would be to apply a combination of molecular dynamics with \emph{ab initio} spin dynamics~\cite{Antropov1996} or with DMFT. However, at present such calculations are hardly feasible in practice. 

Within our approach we describe the paramagnetic state of a system within the disordered local moment picture. In this approach, local moments exist at each magnetic site of a system (in our case, at Cr sites in CrN) and are commonly thought to fluctuate fairly independently. Thoughtful discussions of the DLM model can be found in Ref.s.~\cite{Gyorffy1985-JPhysF-15-1337,Hubbard1979, Hubbard1981, Hasegawa1979, Hasegawa1980}. The status of the DLM approach in the many-body lattice models like the Hubbard or s-f exchange ``Kondo lattice'' model is discussed in Ref.~\cite{Niklasson2003}, where it is argued that though in a complete theory the charge and spin fields are dynamically fluctuating both in space and time, a ``static'' DLM approximation, where one neglects the dynamics of the fluctuations captures an important part of the correlations. In the DLM a correlated system is described in terms of a pseudo-alloy of spin up and spin down components. Combined with the coherent potential approximation (CPA)~\cite{Gyorffy1985-JPhysF-15-1337} it becomes equivalent to the ``Hubbard III'' approximation~\cite{Hubbard1964} for the original many-body problem, and it is used with success in many applications. 

However, the CPA is applicable for the description of a substitutionally disordered system with atoms at the sites of an ideal underlying crystal lattice~\cite{Ruban2008REV}, and therefore cannot be used for treatment of lattice dynamics at finite temperatures. In a previous work\cite{Alling2010-PRB-82-184430}, we took one step towards the simultaneous modeling of magnetic and vibrational finite temperature effects by suggesting two alternative supercell implementations of the DLM calculations. In the first, a specific collinear distributions of up and down magnetic moments arranged to minimize the spin correlation functions were used, in line with the special quasirandom structure (SQS)\cite{Zunger1990-PRL-65-353} methodology. In the second, a magnetic sampling method (MSM) was proposed. In the MSM, the energies of a number of randomly generated magnetic distributions were calculated and their running average was taken as the potential energy of the paramagnetic sample. In Ref.~\cite{Alling2010-PRB-82-184430} it was shown that for CrN MSM calculations are converged already for 40 different magnetic distributions, and the two approaches, the SQS and MSM give almost identical results.  

The SQS approach makes use of the fact that in a static picture with all atoms fixed on ideal lattice points, the description of a spatial disorder between the local moment orientations is a good approximation to model the energetics of the combined space and time fluctuations of magnetic moments in a real paramagnet. Unfortunately, if the vibrations of atoms are to be included, one needs to go beyond the fixed magnetic state described by the SQS. The reason is that if a magnetic state is fixed in time one would see artificial static displacements of atoms off their lattice sites due to forces between the atoms with different orientations of their local moments and with different local magnetic environments. In the CrN case those are likely to be quite large due to the magnetic stress discussed in Ref.~\cite{Fillipetti-PRL-85-5166}. In a real paramagnet, due to the time fluctuations of the local moments, this effects should be at least partially averaged out and suppressed depending on the time scales of the spin fluctuations and atomic motions. 

The MSM could in principle be used to obtain the adiabatic approximation where the magnetic fluctuations are considered to be instantaneous on the time scales of atomic motions. This approximation would be obtained if the forces acting on each atom were averaged over a sufficient number of different magnetic samples during each time step of a molecular dynamics simulation. The obvious drawback in this approach is that a large number of calculations needs to be run in parallel leading to an increase, a factor 40 in our case, in computational demands. Furthermore, as stated above it is not at all clear that this adiabatic approximation is motivated in any system. However, the MSM gives us a very good starting point for the implementation of the DLM picture in a MD framework.

\subsection{Disordered local moments molecular dynamics}
In this work we introduce a method for molecular dynamics simulations of paramagnetic materials within the traditional \emph{ab-initio} MD framework. Starting from the DLM idea of a spatial disorder of local moments, we also change the magnetic state periodically and in a stochastic manner during our MD simulation. In this way we deal with a magnetic state that does not show order either on the length scales of our supercell, or time scale of our simulation. We make an approximation that the magnetic state of the system is completely randomly rearranged with a time step given by a spin flip time ($\Delta t_{sf}$), and with a constraint that the net magnetization of the system should be zero. 
Hence to simulate a paramagnetic system with a spin flip time $\Delta t_{sf}$, we initialize our calculations by setting up a supercell where collinear local moments are randomly oriented and the total moment of the supercell is zero, and run collinear spin-polarized MD for the number of MD time steps ($\Delta t_{MD}$) corresponding to the spin flip time, that is for $\Delta t_{sf}/\Delta t_{MD}$ time steps. Thereafter the spin state is randomized again, while the lattice positions and velocities are unchanged, and the simulation run continues.

Here it is worth to point out that besides the treatment of the many-body effects important for the description of the paramagnetic state at the DLM-LSDA level, or as will be discussed below for the present case: DLM-LSDA+U, we introduce several additional approximations. In particular, we neglect effects due to non-collinear orientations of the local magnetic moments. This, however, is justified for the paramagnetic state well above the magnetic transition temperature~\cite{Gyorffy1985-JPhysF-15-1337}. Note also that magnitudes of the local magnetic moments are allowed to vary as dictated by the self-consistent solution of the electronic structure problem at each step of MD simulation. At the same time, we substitute the true spin dynamics with instantaneous modification of the sample magnetic structure with time steps $\Delta t_{sf}$. Here we follow Ref.~\cite{Gyorffy1985-JPhysF-15-1337} and make use of the physical picture that the simulated system, although ergodic, does not cover its phase space uniformly in time. In the DLM model one assumes that it gets stuck for long times, of the order of $\Delta t_{sf}$, near points characterized by a finite moment at every site pointing in more or less random directions and then moves rapidly (in our case instantly) to another similar point. The states of temporarily broken ergodicity are characterized by classical unit vectors, $\bf e_i$, assigned to each site i and giving the direction of the magnetization averaged over the spatial extent of the i-th site in the supercell and the time $\Delta t_{sf}$. The motion of temporarily broken ergodicity is mainly characterized by changes in the orientational configuration of the moments. 

Note that in Ref.~\cite{Gyorffy1985-JPhysF-15-1337} the magnetic degree of freedom was related to an inverse spin-wave frequency $t_{sw} \sim 1/\omega_{sw} \sim$ 100 fs, which represents the dominating magnetic excitation at low temperatures. However, in the paramagnetic state at high temperatures the relevant magnetic excitations are associated with spin-flips rather than with spin waves. Thus the relevant time scale is better characterized by the spin decoherence time $\Delta t_{dc}$ rather than by the inverse spin-wave frequency. As we pointed out above, the latter was estimated to be of the order 20-50 fs in bcc Fe above $T_C$~\cite{Hellsvik2008}. For CrN, with a $T_N$ around room temperature and probably with weaker exchange interactions, we expect that $t_{dc}$ could be somewhat larger.

However, our procedure makes it possible to model a paramagnetic system for any particular time scale of the spin dynamics. In fact one can span the whole range between the two adiabatic approximations: from the frozen magnetic structure to magnetic configurations that rearrange instantaneously on the time scales of each atomic motion during the MD run. Of course, the appropriate value of this parameter needs to be found with real spin dynamics calculations or taken from experiments. In this paper, we study a range of different spin flip times and their consequences for the obtained structural and thermodynamic properties of CrN. 

\subsection{Calculational details}
All our first-principles calculations in this work is performed using the projector augmented wave method~\cite{Blochl:1994p1407}  as implemented in the Vienna Ab-Initio Simulation Package (VASP)~\cite{Kresse:1993p4005,Kresse:1996p4007,Kresse:1996p4006}. The electronic exchange-correlation effects are modeled using a combination of the Local Density Approximation with a Hubbard Coloumb term (LDA+U)~\cite{Anisimov1991-PRB-44-943} using the double-counting correction scheme suggested by Dudarev \emph{et al.}~\cite{Dudarev1998-PRB-57-1505}. The value of the effective $U$ ($U^{eff}=U-J$) applied only to the Cr $3d$-orbitals is taken as 3 eV, found to be suitable in comparison with several experimentally measured structural and electronic properties in Ref.~\cite{Alling2010-PRB-82-184430}.

Our simulation box, both in the simulation of the cubic and orthorhombic phases, contains 32 Cr and 32 N atoms arranged in a supercell of 2x2x2 conventional cubic unit cells. In the orthorhombic case, the primitive vectors of the supercell are tilted and scaled in line with the results of a structural optimization of this low temperature antiferromagnetic phase.  

The plane wave energy cut-off was set to 400 eV. We used a Monkhorst-Pack scheme~\cite{Monkhorst1976-PRB-13-5188} for sampling of the Brillouin zone using a grid of 2x2x2 k-points. To check the accuracy of the potential energies and pressures a selection of configurations are chosen out of the MD simulation run and recalculated with a higher accuracy. The error arising from for the k-point sampling is relatively constant with a shift of about 35 meV and a standard deviation of less then 2 meV. Hence the relative potential energies that are calculated have a high accuracy. The pressures also have a small constant shift of about 0.2 GPa with a standard deviation less then 0.1 GPa when the electronic structure calculations are converged with respect to the k-point mesh. 

In order to control the temperature of the simulation, avoid artificial energy drift and to minimize the influence of the particular choice of initial magnetic and lattice configurations in the simulations we use a Nose thermostat\cite{Nose1991-PTP-103-1}. The values of the bulk modulus, $K_0$, have been determined by fitting our calculated pressure and volume data to the Birch-Murnaghan equation of state\cite{Birch1947-PhysRev-71-809,Murnaghan1944-PNAS-30-244}.

\begin{equation}
\label{eqn:bme}
P=3K_{0} f_{E} (1+2f_{E})^{5/2}\left(1+2/3(K_0^{'}-4)f_{E}\right)
\end{equation}

\noindent where $K_{0}$ is the bulk modulus and $K_0^{'}$ is the derivative of $K_{0}$ with respect to pressure.
The fulerian strain, $f_E$ is defined as
$f_E=1/2[(V_0/V)^{2/3}-1]$,
where $V$ and $V_0$ are the volume respectively equilibrium volume.

\section{Application to C\lowercase{r}N}
\subsection{The potential energy}
\begin{figure}[htb]
 	\centering
  	\includegraphics[scale=1]{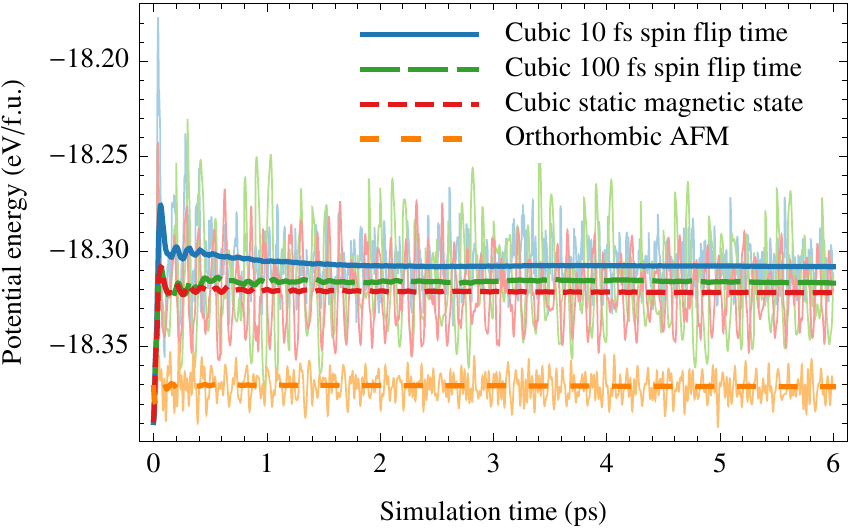}
  	\caption{ Potential energy of cubic paramagnetic CrN as a function of simulation time
  	calculated at 300 K using DLM-MD method. Shown are the results obtained with a spin flip time of 10 fs and 100 fs, as well as with a static magnetic state. Results for conventional AIMD simulations
	carried out for CrN in the orthorhombic antiferromagnetic ground state are also shown for comparison.
  	The potential energy is stable and well converged as can be seen 
  	by the included running averages.}
 	\label{fig:Epot}
\end{figure}

From the MD calculation we extract the potential energies of CrN. As can be seen in Fig.~\ref{fig:Epot}, the potential energy of the system is well conserved. 
To investigate the influence of the spin flip time, the potential energy of CrN is calculated for several $\Delta t_{sf}$. The results are shown in Fig.~\ref{fig:Epot}.

\begin{figure}[htb]
 	\centering
  	\includegraphics[scale=1]{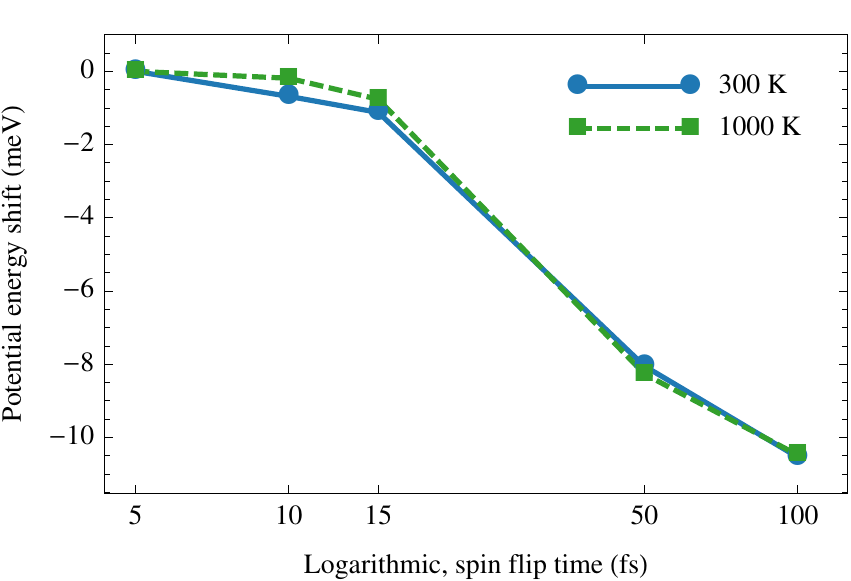}
  	\caption{ Potential energy shift calculated for paramagnetic CrN as a function of the spin flip time $\Delta t_{sf}$. 
  	The shortest spin flip time of 5 fs is taken as reference.}
 	\label{fig:Eshift}
\end{figure}

 In Fig.~\ref{fig:Eshift} these potential energies are collected and shown relative to the potential energy of the calculations with shortest $\Delta t_{sf}$, 5 fs. There is a clear shift in potential energy of about 10 meV from the simulations with the shortest spin flip times of 5 fs to the longest of 100 fs. This can be compared with the total energy reduction due to static relaxations of 15 meV that we get by using the SQS approach treating the magnetic state as frozen in time. We note that for the lower values of $\Delta t_{sf}$, corresponding to fast spin decoherence, there is a plateau were the potential energy is only weakly influenced by $\Delta t_{sf}$. However, between spin flip times of 15 to 50 fs, 
 there is a considerable change in potential energy. Of course, the energy scale should be material specific. We suggest, as a quick test of the importance to consider this effect, a calculation of relaxation energies of a paramagnetic system using the SQS approach~\cite{Alling2010-PRB-82-184430} with a fixed magnetic state through the relaxation. The obtained relaxation energy should correspond to an upper limit on the potential energy dependence on $\Delta t_{sf}$.

\subsection{Pair distances}
\begin{figure}[htb]
 	\centering
  	\includegraphics[scale=1]{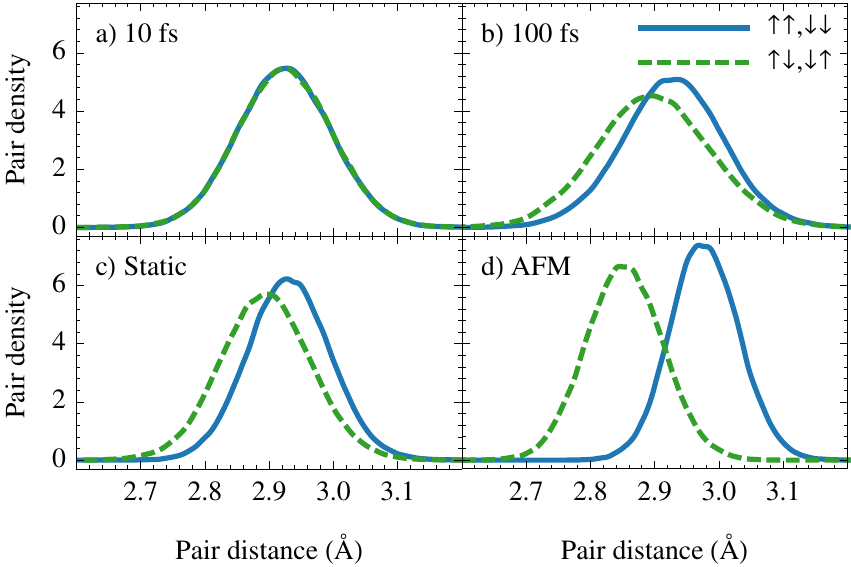}
  	\caption{ Histogram of the Cr - Cr nearest neighbor distances for Cr atoms with parallel (solid line) and antiparallel (dashed line) orientations of local magnetic moments obtained from DLM-MD simulation for the cubic paramagnetic phase at 300 K. 
  	a) $\Delta t_{sf}=$10 fs.
  	b) $\Delta t_{sf}=$100 fs.
  	seen.  
  	c) Static magnetic state.
  	d) The orthorhombic antiferromagnetic phase of CrN calculated with conventional AIMD for comparison.
	}
 	\label{fig:Pairdist}
\end{figure}
In order to analyze the difference between the proposed DLM-MD simulations and magneto-static MD in more details, an investigation of the local environment of the different atoms is carried out, especially the Cr - Cr metal nearest neighbor distances. In Fig.~\ref{fig:Pairdist} histograms are shown of all the Cr - Cr nearest neighbor distances. These are also separated into $\uparrow\uparrow,\downarrow\downarrow$ and $\uparrow\downarrow,\downarrow\uparrow$ pairs. Hence we can see the effect of the magnetic state on the distribution of pair distances. In Fig.~\ref{fig:Pairdist}a the $\Delta t_{sf}$ is very short, 10 fs, hence the atoms do not have time to adjust their positions for the current orientation of local magnetic moments and we do not see any difference in distances between the $\uparrow\uparrow,\downarrow\downarrow$ and the $\uparrow\downarrow,\downarrow\uparrow$ pairs. In Fig.~\ref{fig:Pairdist}b, the spin flip time is increased to 100 fs and now the atoms have had sufficient time to move towards the energetically preferential positions. Consequently, a shift in pair distances between $\uparrow\uparrow,\downarrow\downarrow$ and the $\uparrow\downarrow,\downarrow\uparrow$ is evident. Fig.~\ref{fig:Pairdist}c is obtained with the same orientation of local moments during the whole MD run. Here we also see a splitting in the pair distances between the $\uparrow\uparrow,\downarrow\downarrow$ and the $\uparrow\downarrow,\downarrow\uparrow$ pairs, which is of the same order as in the previous case.  Hence, 100 fs between the re-arrangement of magnetic configurations is long enough for the atomic nuclei to adjust considerably their positions in the supercell to the given magnetic configuration. 
In the last figure, Fig.~\ref{fig:Pairdist}d, the pair distances are shown for the low temperature antiferromagnetic orthorhombic ground state for comparison. Here the $\uparrow\uparrow,\downarrow\downarrow$  and $\uparrow\downarrow,\downarrow\uparrow$ pairs of magnetic moments are arranged in an ordered way, see e.g. Fig. 4 in Ref.~\cite{Alling2010-PRB-82-184430}, that allows for maximal relaxation of atomic coordinates in combination with a structural relaxation of the unit cell, giving rise to a large separation between the two different kinds of pairs.

A possibility of statistical correlations between the atomic distances and the orientation of atomic moments also in a dynamically changing paramagnetic phase is indeed an intriguing thought experiment. Although we can not rule out its existence from principal considerations, to the best of our knowledge it has never been reported in experiments. However, we note that our two main approximations in the present DLM-MD, the usage of collinear moments and the temporarily broken ergodicity of the DLM approach, are likely to introduce inaccuracies that exaggerate those local spin-lattice correlations when a slow spin dynamics is modelled. Therefore, when the here suggested method is used, a smaller value of $\Delta t_{sf}$, corresponding to the absence of differences in distances between atoms with parallel and antiparallel local moments, like in Fig.~\ref{fig:Pairdist}a, should be recommended. The presence of an energy plateau in Fig.~\ref{fig:Eshift} seems to indicate that this should be a reasonable approach.

\subsection{Bulk modulus of C\lowercase{r}N}
\begin{figure}[htb]
 	\centering
  	\includegraphics[scale=1]{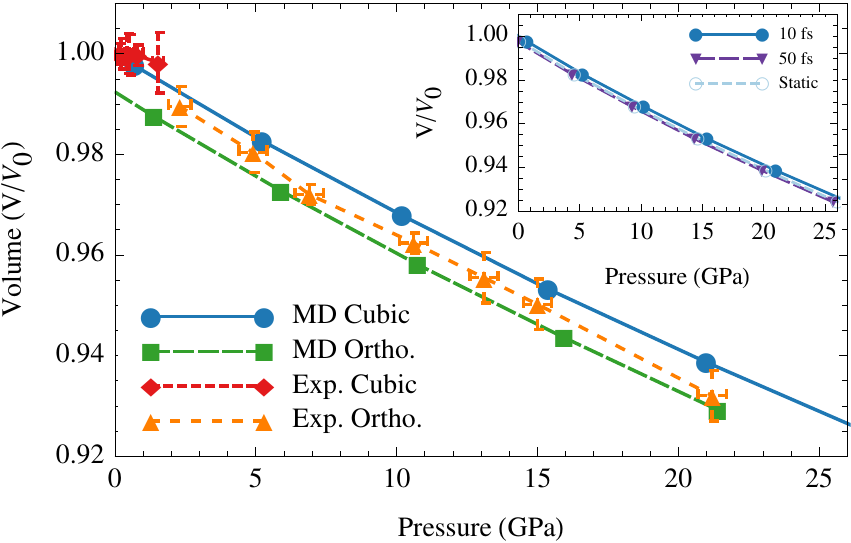}
  	\caption{ 
	Volume as a function of pressure for the cubic and orthorhombic phase from MD simulations at 300 K.
	The equation of state for the cubic phase is calculated using a  spin flip time of 10 fs. The calculated volumes
	are normalized with the calculated equilibrium volume of the cubic phase, 
	and the experimental points\cite{Rivadulla2009-Nat.Mater-8-947} with the measured equilibrium volume of the cubic phase.
	The inset shows the dependence 
	on spin flip time for the calculated equation of state	}
 	\label{fig:EOS}
\end{figure}
Our main goal with this work is to study the equation of states, and in particular the bulk modulus of paramagnetic CrN which has recently been discussed in the literature~\cite{Rivadulla2009-Nat.Mater-8-947, Alling2010-NatMater-9-283}. Using our DLM-MD approach we are able to calculate volume as a function of temperature and pressure for both the paramagnetic cubic and the antiferromagnetic orthorhombic phases. In the former case we also investigate if there is an impact of the value of the spin flip time parameter on the equation of state. Thus we are able to investigate how both the dynamical change of magnetic configurations in the paramagnetic state and the lattice vibrations, neglected in previous theoretical works but of course present in the experiments, impact on the compressibility. Figure~\ref{fig:EOS} shows the calculated volume as a function of pressure for the two phases at 300 K and compare them to the experimental measurements by Rivadulla~\emph{et al.}~\cite{Rivadulla2009-Nat.Mater-8-947}. One sees very good agreement between theoretical and experimental equations of state. In particular the relative shift in volumes between the two phases is reproduced within the measured error bars. The calculated slope of the orthorhombic phase agrees well with the measured values for this phase where the measurement is done over a large pressure range. The inset in Fig.~\ref{fig:EOS} shows the influence of the spin flip time on the volume versus pressure curves in paramagnetic cubic CrN. A change in $\Delta t_{sf}$ introduces a small shift of the volumes, but does not influence the slope of the curves. 

Our results confirm the possibility of a pressure induced phase transition from the cubic paramagnetic to the orthorhombic antiferromagnetic phase due to the slightly smaller volume of the latter, in line with previous investigations. Importantly, as can be seen in Table \ref{tbl:K0} and from the slopes of the curves in Fig.~\ref{fig:EOS}, the bulk modulus is found to be very similar between the two phases. This is the case both at 300 K, 1000 K, and in the static 0 K calculations. The calculations of the orthorhombic low temperature phase at 1000 K is of course not of relevance for any comparison with experiments, but is included to show with certainty that temperature induced vibrations is not influencing the \emph{difference} in bulk modulus between the phases. At T=300K and P=0GPa we find $K_{0}^{para}=290$~GPa while $K_{0}^{AFM}=286$~GPa. This gives an insignificant difference of 4 GPa, far from the collapse of 25\% or 85 GPa suggested in Ref.~\cite{Rivadulla2009-Nat.Mater-8-947} to follow the transition from cubic to orthorhombic structures. A variation of the time between the rearrangement of the magnetic configurations do not influence the value of the bulk modulus in any appreciable way. Thus, explicit considerations of temperature induced magnetic fluctuations and lattice vibrations do not change the main conclusions from previous works~\cite{Alling2010-NatMater-9-283, Alling2010-PRB-82-184430}: There is no theoretical support for a collapse of the bulk modulus of CrN upon the pressure induced phase transition. 

\begin{table}
\caption{\label{tbl:K0} Calculated bulk modulus, $K_{0}$, of CrN in the orthorhombic antiferromagnetic and the 
	cubic paramagnetic phases obtained at ambient pressure and 0, 300, and 1000 K respectively.
	}
\begin{ruledtabular}
\begin{tabular}{lccc}
 \text{Structure} & \text{Static 0K} & \text{MD 300 K} & \text{MD 1000 K} \\
 \hline
 \text{Orthorhombic AFM} & 290 & 286 & 261 \\
 \text{Cubic Paramagnetc} & 299 & 290 & 269
\end{tabular} 
\end{ruledtabular}
\end{table}

\section{Summary}
We present a method for calculation of thermodynamic properties of magnetic materials in their high temperature paramagnetic state. 
We use \emph{ab-initio} molecular dynamics and simulate the paramagnetic state with disordered non-vanishing local magnetic moments. Random configurations of the local moments in the simulation cell are switched at predetermined time intervals. Hence we can capture the influence of the dynamically disordered magnetic state on the lattice dynamics as it develops during the simulation. We apply this method to CrN which is known to have a strong interaction between the magnetic state and the lattice. We find that there is a connection between how fast the local moments are allowed to flip and the calculated potential energy. If the spin flip time is short, $\sim10$ fs, the lattice do not have time to respond, but if the spin flip time is increased to about 100 fs then the atomic positions start to show clear relaxation effects. We apply this disordered local moments molecular dynamics method to the calculation of the equation of state of paramagnetic cubic CrN and compare with calculations for the orthorhombic antiferromagnetic phase and with experiments. In particular we calculate the debated bulk modulus and find that there is only a very small difference, and definitely no collapse, in $K_0$ between the orthorhombic antiferromagnetic phase and the cubic paramagnetic phase.

\begin{acknowledgments}
	We gratefully acknowledge financial support by the Swedish Foundation for Strategic Research (SSF) Program in Materials Science for Nanoscale Surface Engineering, MS$^2$E, the Swedish Research Council (VR), and G\"oran Gustafsson foundation for research in Natural Sciences and Medicine. . The simulations were carried out using supercomputer resources provided by the Swedish national infrastructure for computing (SNIC).
\end{acknowledgments}


%

\end{document}